\documentclass[amsmath,amssymb,superscriptaddress,prd,nofootinbib,twocolumn]{revtex4-2}
\usepackage{graphicx}
\usepackage{amsmath}
\usepackage{amssymb}
\usepackage{amsfonts}
\usepackage{color}
\usepackage{dsfont}
\usepackage{tikz}
\usepackage{ragged2e}
\usetikzlibrary{calc,decorations.markings}
\usepackage[colorlinks=true,linkcolor=blue,citecolor=blue,urlcolor=blue]{hyperref}
\usepackage{physics}
\usepackage{soul}
\usepackage{float}
\usepackage{subfig}

\newcommand{\x}{\mathsf{x}}

\usetikzlibrary{decorations.pathmorphing, patterns,shapes}
\definecolor{lime}{HTML}{A6CE39}
\DeclareRobustCommand{\orcidicon}{%
	\begin{tikzpicture}
	\draw[lime, fill=lime] (0,0) 
	circle [radius=0.16] 
	node[white] {{\fontfamily{qag}\selectfont \tiny ID}};
	\draw[white, fill=white] (-0.0625,0.095) 
	circle [radius=0.007];
	\end{tikzpicture}
	\hspace{-2mm}
}
\foreach \x in {A, ..., Z}{%
	\expandafter\xdef\csname orcid\x\endcsname{\noexpand\href{https://orcid.org/\csname orcidauthor\x\endcsname}{\noexpand\orcidicon}}
}

\begin{document}
\title{Thermodynamic uncertainty relations for relativistic quantum thermal machines}
\author{Dimitris Moustos\orcidA{}}
\email{dimitris.moustos@newcastle.ac.uk}
\affiliation{School of Mathematics, Statistics, and Physics, Newcastle University, Newcastle upon Tyne NE1 7RU, United Kingdom}
\author{Obinna Abah\orcidB{}}
\email{obinna.abah@newcastle.ac.uk}
\affiliation{School of Mathematics, Statistics, and Physics, Newcastle University, Newcastle upon Tyne NE1 7RU, United Kingdom}
\date{\today}

\begin{abstract}
We investigate a two-qubit SWAP thermal machine -- a streamlined analogue of the four-stroke Otto cycle -- whose working medium comprises inertially moving Unruh-DeWitt qubit detectors, each coupled to a thermal quantum field bath prepared at a different temperature. In the presence of relative motion between the working medium and the thermal baths, we derive thermodynamic uncertainty relations (TURs) that quantify the trade-off between performance, entropy production, and power fluctuations. Our analysis identifies regimes where relativistic motion leads to stronger violation of classical TURs, previously observed in static quantum setups. In addition, we establish generalized performance bounds for the thermal machine operating as either a heat engine or a refrigerator, and discuss how relativistic motion can enhance their performances beyond the standard Carnot limits defined by rest-frame temperatures.

\end{abstract}

\maketitle

%===========================================================================================================================
\section{Introduction}
Thermal machines have played a central role in the advancement of society, from the advent of the industrial revolution to the development of modern technologies. 
Among the most fundamental examples of thermal machines operating cyclically between a hot reservoir at temperature $T_h$ and a cold reservoir at temperature $T_c$ are heat engines and refrigerators.  A heat engine converts thermal energy into mechanical work, whereas a refrigerator consumes work to extract heat from a cold reservoir.  The second law of thermodynamics imposes a universal upper bound on the efficiency of any heat engine  defined by the Carnot efficiency, $\eta_C=(T_h-T_c)/T_h$, while the maximum coefficient of performance (COP) for a refrigerator is $\varepsilon_C=T_c/(T_h-T_c)$ \cite{Callen}.

Advancements in nanoscale device fabrication and the coherent control of atoms have paved the way for the realization of thermal machines operating in the quantum regime, 
enabling the investigation into the thermodynamic behavior of quantum systems and the role of quantum information in thermodynamic processes \cite{vinjanampathy2016quantum,goold2016role,binder2019thermodynamics,Deffner:Campell,therm:eng:rev,potts2024quantumthermodynamics}. At the microscopic scale, stochastic fluctuations of thermodynamic variables and irreversible entropy production become inherently significant, often affecting the performance of quantum thermal machines. Consequently, elucidating the fundamental principles that govern nonequilibrium thermodynamic processes is of crucial importance. In this context, various symmetry relations, known as fluctuation theorems \cite{RevTPM,EFT,Seifert_2012,EspositoRev,Jarz,FT:Seifert,Crooks}, have been formulated to characterize the probability distributions of stochastically fluctuating quantities. They provide an extension of the second law of thermodynamics to the framework of stochastic thermodynamics, quantifying the occurrence of negative fluctuations of entropy.

Recently,  the formulation of the trade-off between performance, 
entropy production and relative fluctuations of output power, in terms of a thermodynamic uncertainty relation (TUR) \cite{Barato2015,Barato:Seifert:16,TUR:Horowitz}, has been established. The TUR connects the noise-to-signal ratio of a thermodynamic current (e.g., heat, work, or particle number) and dissipation. Originally derived for classical Markovian stochastic processes, it reads \cite{Barato2015}:
\begin{align}\label{TUR}
    \frac{\text{Var}[\mathcal{J}]}{\expval{\mathcal{J}}^2}\geq\frac{2}{\expval{\Sigma}},
\end{align}
where $\expval{\mathcal{J}}$ is the steady-state mean current, $\text{Var}[\mathcal{J}]\!=\!\expval{\mathcal{J}^2}-\expval{\mathcal{J}}^2$ denotes its fluctuation,  and $\expval{\Sigma}$ is the average entropy production. This relation reveals a fundamental trade-off: achieving higher precision in a current---i.e., reducing its fluctuations---entails increased dissipation, quantified by the entropy production. For classical heat engines operating in a nonequilibrium steady state, the TUR implies a corresponding trade-off between the mean output power $\expval{P}$,
its fluctuations $\text{Var}[P]$, and  the efficiency, $\eta$, expressed as \cite{Pietzonka2018}:
\begin{align}
    \eta\leq\frac{\eta_C}{1+2\expval{P}T_c/\text{Var}[P]},
\end{align}
which indicates that approaching the Carnot limit at finite power output is possible only at the cost of diverging power fluctuations. While the TUR holds under broad conditions for classical stochastic processes, it can be violated in the quantum regime, where such violations have, in some cases, been linked to potential enhancements in the efficiency of quantum heat engines (see, e.g.,  \cite{Segal2018,Krzysztof2018,qTUR,Liu:Segal,Saryal:Segal, TUR:marti,Menczel_2021,PP3lvl}). 

In recent years, there has been growing interest in understanding how relativistic phenomena influence the performance of quantum thermal devices \cite{Munoz2012,Pena2016,arias2018unruh,myers2021quantum}. Significant focus has been placed on thermal machines whose working medium consists of Unruh-DeWitt (UDW) detectors \cite{Unruh,DeWitt,Birrell:Davies}---qubits coupled to quantum fields while following arbitrary worldlines in a background spacetime. 
A notable example is the Unruh Otto heat engine \cite{arias2018unruh,gray2018,XU2020135201,Unruh:entangl,barman2022,Mukherjee_2022}: during the isochoric steps of the Otto cycle, the detector undergoes uniform acceleration while
interacting with a quantum field in its vaccum state, which, due to the
Unruh effect \cite{Unruh}, effectively acts as a thermal reservoir. This framework has been further generalized to incorporate different uniformly accelerated trajectories, such as circular motion, as well as alternative working media such as qutrit detectors, and scenarios involving instantaneous interactions between the detector and the field (see, e.g. \cite{npapadatos,gallock2023quantum,NK:DM,gallock2024relativistic,HT25,sarkar2025}). In addition, studies have explored how curved spacetimes, for instance black hole geometries, affect the performance of quantum thermal machines \cite{DEB,Ferketic:Deffner,kollas2024,misra2024black,DM:OA}.  

On the other hand, a stochastic thermodynamic framework has been developed to analyze the distributions of heat and work, as well as to derive various forms of fluctuation theorems for systems such as particles at relativistic energies \cite{Pal2020,Paraguass2021}, and  quantum fields in flat or curved spacetime \cite{QFT:FT,QFT:work,QFT:FT:EMM,costa2025workdistributionquantumfields,FT:covariant,FTcurved}. However, the effects of relative motion between the working medium and the thermal bath on fluctuation theorems and TURs remain largely unexplored. In this work, we derive fluctuation relations and TURs in the context of a quantum thermal machine whose working medium consists of inertially moving UDW qubit detectors.

An UDW detector moving at a constant relativistic speed experiences a frequency-dependent effective temperature, which can be either hotter or colder than the ambient temperature of the thermal field bath. Building on this observation, we previously introduced a quantum SWAP heat engine (a two-stroke engine) \cite{DMOA2025surpassingcarnot,Nori2007,campisi2015,TUR:SWAP,Herrera2023,SWAP:null}, where the two-qubit working medium is modeled by inertially moving UDW detectors, each locally coupled to a thermal field bath at a different temperature. Here, we derive generalized efficiency bounds that govern the performance of the thermal machine operating either as a heat engine or as a refrigerator. These bounds arise from the second law of stochastic thermodynamics. We identify regimes in which relativistic motion leads to stronger violations of the classical TUR bound, previously observed in the quantum regime.

The paper is organized as follows. In Sec. \ref{Sec:UDW}, we briefly review the UDW detector model and define the effective temperatures experienced by moving detectors. In Sec. \ref{Sec:SWAP:mov}, we introduce a two-qubit SWAP engine with moving detectors as the working medium and derive fluctuation theorems governing the stochastic dynamics of work and heat distributions. In this setup, Sec. \ref{sec:TUR} derives a thermodynamic uncertainty relation, and Sec. \ref{Sec:eff:bounds} provides bounds on the efficiency and COP of the thermal machine. Finally, Sec. \ref{Sec:conclusions} summarizes and discusses our results. Throughout, we use natural units, setting $\hbar=c=k_B=1$.

%===========================================================================================================================
\section{Inertially moving UDW detector through a thermal bath}\label{Sec:UDW}

We consider an UDW detector  \cite{Unruh,DeWitt,Birrell:Davies} on a worldline $\x(\tau)$, parametrized by proper time $\tau$, in a (3+1)-dimensional Minkowski spacetime. We model the detector as a pointlike two-level quantum system (qubit) with ground state $\ket{g}$ and excited state $\ket{e}$, separated by an energy gap $\omega$. The detector's free Hamiltonian is $H_D=\omega\sigma_z/2$, 
where $\sigma_z$ is the standard Pauli-z matrix. We allow the detector to interact with a massless scalar quantum field $\phi$, initially prepared in a state $\rho_{\phi}$. In the interaction picture, the interaction Hamiltonian reads
\begin{align}
    H_{\text{int}}(\tau)=\lambda \mu(\tau)\phi(\mathsf{x}(\tau)),
\end{align}
where $\lambda$ is a coupling constant, $\mu(\tau)=e^{i\omega\tau}\sigma_++e^{-i\omega\tau}\sigma_-$ is the detector's monopole moment operator expressed in
terms of the ladder operators $\sigma_+=\ket{e}\bra{g}$ and $\sigma_-=\ket{g}\bra{e}$, and $\phi(\mathsf{x}(\tau))$ is the field evaluated on the detector’s worldline. 

We next  specialize to the case where both the detector's trajectory and field's state are stationary. In this case, the pullback of the Wightman two-point correlation function, $\mathcal{W}(\tau,\tau'):=\expval{\phi(\x(\tau))\phi(\x(\tau'))}_{\rho_{\phi}}$, along the detector's worldline depends only on the proper time deference $\tau-\tau'$ between any two points on the detectors' worldline---i.e., $\mathcal{W}(\tau,\tau')=\mathcal{W}(\tau-\tau')$ \cite{Birrell:Davies}. The detector’s transition rate \cite{Birrell:Davies}---the probability per unit of proper time for a transition between its energy levels---is given by the Fourier transform of the Wightman function, 
\begin{align}
\mathcal{G} (\omega):=\int_{-\infty}^{+\infty}ds\,e^{-i\omega s}\mathcal{W}(s).
\end{align}

In the weak-coupling interaction limit, the detector reaches at late times a (non-equilibrium) steady state, described by the reduced density matrix \cite{BJA:DM}:
\begin{align}\label{app:dm}
    \rho_D=\frac{e^{-\beta^{\text{eff}}(\omega)H_D}}{\text{tr}(e^{-\beta^{\text{eff}(\omega)}H_D})},
\end{align}
where the frequency-dependent effective temperature $T^{\text{eff}}(\omega)=1/\beta^{\text{eff}}(\omega)$
is defined through the detailed balance condition, $\mathcal{G}(-\omega)=e^{\omega/T^{\text{eff}}(\omega)}\mathcal{G}(\omega)$, which relates the detector’s excitation and de-excitation rates (see, eg., \cite{BJA:DM,Effective:Unruh,Biermann2020,bunney2023circular,bunney2024ambient,Parry2025}). Equivalently,
\begin{align}\label{Effective:Temperature}
    T^{\text{eff}}(\omega):=\omega\bigg/\ln\left(\frac{\mathcal{G}(-\omega)}{\mathcal{G}(\omega)}\right).
\end{align}

We now turn to the situation in which the field is prepared in a thermal state with inverse temperature $\beta=1/T$, and the detector is assumed to move inertially with a constant velocity $\upsilon$, following the trajectory $\x(\tau)=(\gamma\tau,\gamma\upsilon\tau,0,0)$, where $\gamma=1/\sqrt{1-\upsilon^2}$ is the Lorentz factor. In this case, the detector's transition rate is given by
\cite{matsas95,COSTA1995,Papadatos,verch2025}
\begin{align}\label{transition:rate:eq}
     \mathcal{G} (\omega)=\frac{\lambda^2}{4\pi\beta\gamma\upsilon}\ln\left(\frac{1-e^{-\beta\gamma(1+\upsilon)\omega}}{1-e^{-\beta\gamma(1-\upsilon)\omega}}\right).
\end{align}
The detector then perceives an effective temperature according to \eqref{Effective:Temperature}, which in the high-temperature regime ($\beta\omega\ll1$) is colder than the ambient temperature $T$ of the field bath, while in the low-temperature regime ($\beta\omega\gg1$) it appears hotter. Moreover, in the ultra-relativistic limit $\upsilon\to1$, the transition rate vanishes, and the detector perceives the field as being in its vacuum state. For a more detailed discussion, see Refs. \cite{matsas95,COSTA1995,Papadatos,verch2025}.

%===========================================================================================================================

\section{Otto cycle with moving qubits}\label{Sec:SWAP:mov}
We consider a two-stroke quantum SWAP thermal machine \cite{Nori2007,campisi2015,TUR:SWAP,Herrera2023,SWAP:null}, consisting of two qubits labeled $A$ and $B$, with respective transition frequencies $\omega_A$ and $\omega_B$.  Each qubit is described by the Hamiltonian $H_i=\omega_i\sigma_z^i/2$, with $i\in\{A,B\}$. 
Initially, qubit $A$ is brought into thermal equilibrium with a hot reservoir at temperature $T_A$, while qubit $B$ with a cold reservoir at $T_B$ (that is, $T_A>T_B$). The initial state of the two-qubit system is  
\begin{align}
    \rho_0=\frac{e^{-\beta_AH_A}}{Z_A}\, \otimes\, \frac{e^{-\beta_BH_B}}{Z_B},
\end{align}
where $\beta_i=T_i^{-1}$ and $Z_i\!=\!\text{tr}(-e^{\beta_iH_i})$ are the inverse temperatures and partition functions, respectively. After thermalization, the qubits are decoupled from their reservoirs and allowed to interact via a SWAP unitary,
%\begin{align}
%    U=\begin{pmatrix}
%1 & 0 & 0 & 0\\
%0 & 0 & 1 & 0\\
%0 & 1 & 0 & 0\\
%0 & 0 & 0 & 1
%\end{pmatrix},
%\end{align}
which exchanges their states. Subsequently, each qubit is re-coupled to its thermal reservoir, allowing it to re-thermalize. This completes a two-stroke quantum engine cycle---an analogue of the four-stroke Otto cycle \cite{binder2019thermodynamics}.

Depending on the transition frequencies of the qubits, the engine exhibits three distinct regimes of operation---see \cite{Nori2007,campisi2015,TUR:SWAP,Herrera2023}. Specifically, when $\omega_B/\omega_A<\beta_A/\beta_B$, the device operates as a refrigerator, consuming work ($\expval{W}\!>\!0$) to extract heat from the cold reservoir. In contrast, when $\beta_A/\beta_B<\omega_B/\omega_A<1$, the device functions as a heat engine, absorbing heat from the hot reservoir to produce useful
work ($\expval{W}<0$). Finally, when $\omega_B/\omega_A>1$, the machine acts as a heat accelerator, consuming work to enhance the heat flow from the hot to the cold reservoir.

We now consider that the two-qubit working medium of the engine is represented by two UDW detectors moving at constant speeds $\upsilon_A$ and $\upsilon_B$. As in the standard UDW detector framework (e.g., for uniformly accelerated detectors), the relativistic motion of the detectors is assumed to be externally sustained and their trajectories fixed, without explicitly modelling the associated energetic cost or backreaction. During the first stroke of the cycle, the moving qubits $A$ and $B$ interact with hot and cold thermal field baths at temperatures $T_A$ and $T_B$ (with $T_A>T_B$), respectively. Due to their motion, each qubit  experiences the effective temperatures denoted by $T^{\text{eff}}_A$ and $T^{\text{eff}}_B$. Once both qubits reach their respective steady states $\rho_D$, with $D\in\{A,B\}$, the engine proceeds through the subsequent strokes of the cycle as in the static configuration.

To characterize the performance of the engine, we introduce the random variables $Q_H$ and $W$, and make use of the cumulant generating function  
\begin{align}
    C(\chi_w,\chi_h)=\ln\expval{e^{i\chi_wW+i\chi_hQ_H}},
\end{align}
where $\chi_w$ and $\chi_h$ denote the counting fields associated with work $W$ and heat exchange with the hot reservoir $Q_H$ respectively \cite{Schaller2014,Strasberg2022}.
The cumulants of $W$ and $Q_H$  can then be obtained by differentiating the cumulant generating function with respect to the corresponding counting fields:
\begin{align}
    \langle\langle W^m Q_H^{n}\rangle\rangle=(-i)^{m+n}\frac{\partial^{m+n}C(\chi_w,\chi_h)}{\partial\chi^m_w\partial\chi^n_h}\Bigg|_{\chi_w=\chi_h=0}.
\end{align}
Following the two-point measurement scheme \cite{RevTPM}, which allows for the joint estimation of $W$ and $Q_H$ with  probability distribution $P(W,Q_H)$, the cumulant generating function can be expressed as \cite{TUR:SWAP,Sacchi2021A,Sacchi2021}
\begin{widetext}
\begin{align}
    C(\chi_w,\chi_h)=\ln\bigg\{\!\text{tr}\bigg[ U^\dagger\!\left(e^{i(\chi_w-\chi_h)H_A} e^{i\chi_wH_B}\right)U\left(\!e^{-i(\chi_w-\chi_h)H_A} e^{-i\chi_wH_B}\right)\rho_A\otimes\rho_B\bigg]\bigg\}.
\end{align}
This admits the explicit form
\begin{align}\label{generating:function}
    C(\chi_w,\chi_h)=\ln\left(\frac{\cosh\bigg(\frac{1}{2}(\beta^{\text{eff}}_A\omega_A+i(\omega_A(\chi_w-\chi_h)+\omega_B\chi_w))\bigg)\cosh\bigg(\frac{1}{2}(\beta^{\text{eff}}_B\omega_B-i(\omega_A(\chi_w-\chi_h)+\omega_B\chi_w))\bigg)}{\cosh(\beta^{\text{eff}}_A\omega_A/2)\cosh(\beta^{\text{eff}}_B\omega_B/2)}\right).
\end{align}
\end{widetext}
The first moments of work and input heat are then obtained as
\begin{align}\label{work:average}
    \expval{W}\!=\!\frac{\!\omega_B\!-\!\omega_A\!}{2}\bigg(\!\tanh(\frac{\beta^{\text{eff}}_B\omega_B}{2})\!-\!\tanh(\frac{\beta^{\text{eff}}_A\omega_A}{2})\!\bigg),
\end{align}
\begin{align}\label{heat:average}
    \expval{Q_H}
    \!=\!\frac{\omega_A}{2}\bigg(\!\tanh(\frac{\beta^{\text{eff}}_B\omega_B}{2})\!-\!\tanh(\frac{\beta^{\text{eff}}_A\omega_A}{2})\!\bigg).
\end{align}
Moreover, the entropy production evaluates as \cite{abah2014efficiency,Landi2021}
\begin{align}\label{entropy}
    \expval{\Sigma}&=(\beta^{\text{eff}}_B-\beta^{\text{eff}}_A)\expval{Q_H}+\beta^{\text{eff}}_B\expval{W}.
\end{align}

From the explicit form of the cumulant generating function in Eq. \eqref{generating:function}, it is straightforward to verify the identity $ C(i\beta^{\text{eff}}_B,i(\beta^{\text{eff}}_B-\beta^{\text{eff}}_A))=0$, which leads to
\begin{align}\label{FT}
    \expval{e^{-\Sigma}}=1,
\end{align}
known as the integral fluctuation theorem \cite{Jarz,FT:Seifert}. Combined with the Jensen's inequality, $e^{\expval{x}}\leq\expval{e^x}$, Eq. \eqref{FT} implies the second law of thermodynamics 
\begin{align}\label{2nd:law}
    \expval{\Sigma}\geq0,
\end{align}
that is the average entropy production is always positive. In addition, we can  verify the symmetry condition $C(i\beta^{\text{eff}}_B-\chi_w,i(\beta^{\text{eff}}_B-\beta^{\text{eff}}_A)-\chi_h)=C(\chi_w,\chi_h)$, which directly leads to the exchange fluctuation theorem \cite{RevTPM,EFT,Seifert_2012,EspositoRev,Campisi_2014}
\begin{align}
    \frac{P(Q_H,W)}{P(-Q_H,-W)}=e^{\Sigma},
\end{align}
where $P(-Q_H,-W)$ denotes the probability distribution associated with the time-reversed process.

\section{Thermodynamic uncertainty relations}\label{sec:TUR}

Employing the cumulant generating function, we determine the fluctuations (second cumulants) of the work, $\text{Var}[W]$, and the heat exchanged with the hot bath, $\text{Var}[Q_H]$, respectively,  as
\begin{align}
    \text{Var}[W]=\frac{(\omega_A-\omega_B)^2(2+\cosh(\omega_A\beta^{\text{eff}}_A)+\cosh(\omega_B\beta^{\text{eff}}_B))}{8\cosh^2(\omega_A\beta^{\text{eff}}_A/2)\cosh^2(\omega_B\beta^{\text{eff}}_B/2)},
\end{align}
\begin{align}
    \text{Var}[Q_H]=\frac{\omega_A^2}{4}\left(\text{sech}^2\left(\frac{\omega_A\beta^{\text{eff}}_A}{2}\right)+\text{sech}^2\left(\frac{\omega_B\beta^{\text{eff}}_B}{2}\right)\right),
\end{align}
where $\text{sech}(x)\equiv1/\cosh x$ is the hyperbolic secant. We then evaluate the corresponding noise-to-signal ratios $\text{Var}[Q_H]/\expval{Q_H}^2$ and $\text{Var}[W]/\expval{W}^2$, from which we obtain that
\begin{align}\label{SNR}
    \frac{\text{Var}[Q_H]}{\expval{Q_H}^2}=\frac{\text{Var}[W]}{\expval{W}^2}=\frac{f(\beta^{\text{eff}}_A\omega_A-\beta^{\text{eff}}_B\omega_B)}{\expval{\Sigma}}-1,
\end{align}
where $f(x)=x\coth\left(x\right)$.

We verify numerically that $f(x)\geq2$ is satisfied for all $x$, thereby leading to the thermodynamic uncertainty relation
\begin{align}\label{bound:analytic}
     \frac{\text{Var}[\mathcal{J}]}{\expval{\mathcal{J}}^2}\geq\frac{2}{\expval{\Sigma}}-1,
\end{align}
which implies that the classical TUR can be violated. We note that the uncertainty relation \eqref{bound:analytic} was previously derived for the static qubit engine case in Refs. \cite{Sacchi2021A,Sacchi2021}. Here, we show that the same relation remains valid when the working medium undergoes uniform relativistic motion.

\begin{figure}[tp!]
    \centering
    \subfloat[$\upsilon_A=0,\upsilon_B=0.8$]{\includegraphics[width=0.48\linewidth]{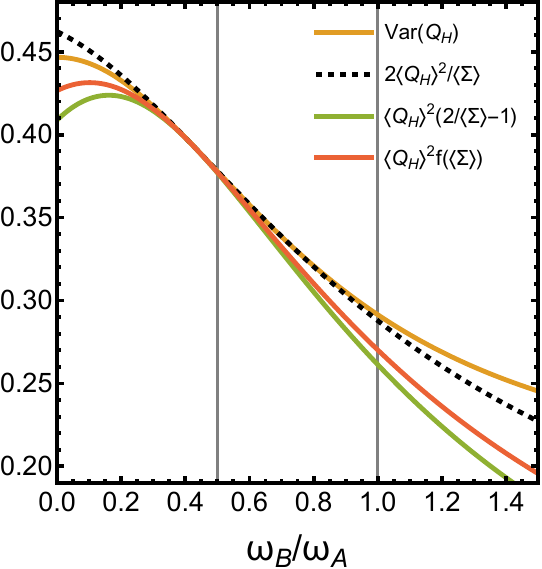}}\hspace{0.18cm}\subfloat[$\upsilon_A=0.8,\upsilon_B=0$]{\includegraphics[width=0.49\linewidth]{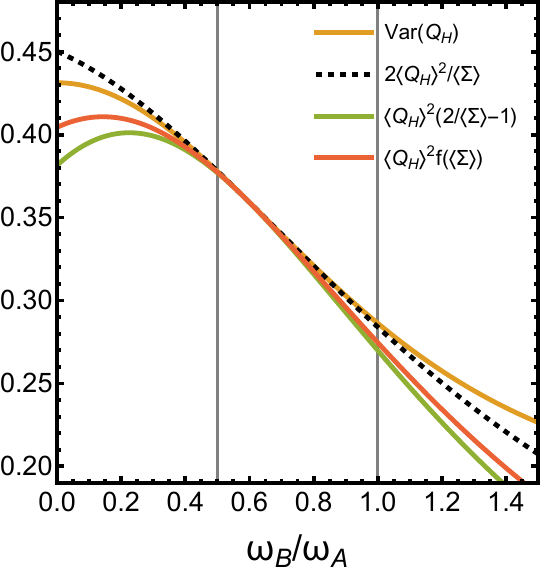}}
    \caption{\justifying  Heat variance, $\text{Var}(Q_H)$, as a function of the transition frequency ration $\omega_B/\omega_A$, compared with the bound \eqref{bound:analytic}, the standard TUR \eqref{TUR}, and the generalized bound \eqref{TUR:gen}. The temperature ratio between the thermal baths is fixed at $\beta_A/\beta_B = 1/2$. (a)  Qubit B moves through the cold bath with speed $\upsilon_A=0.8$. (b) Qubit A moves through the hot bath with speed $\upsilon_A=0.8$. The vertical lines indicate the boundary between the different regimes of operation of the machine.}
    \label{fig:bounds}
\end{figure}
\begin{figure}[tp]
    \centering
    \includegraphics[width=0.48\linewidth]{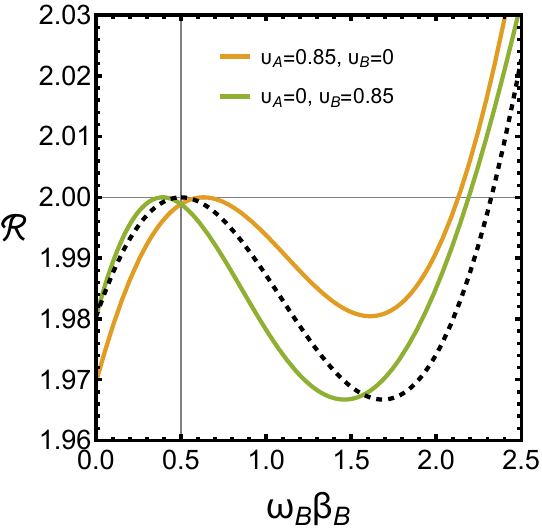}\hspace{0.18cm}
    \includegraphics[width=0.48\linewidth]{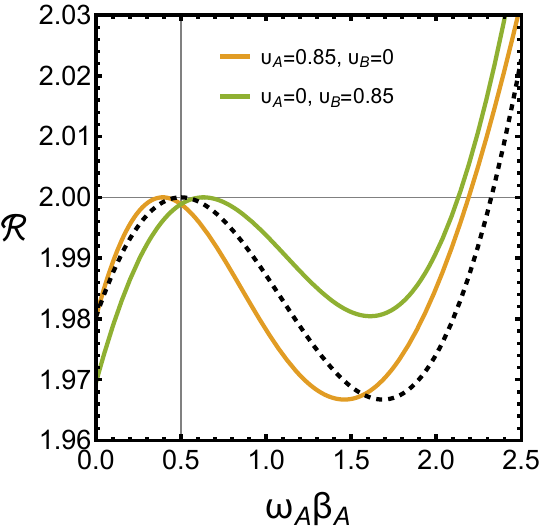}
    \caption{\justifying The ratio
$\mathcal{R}$ for different qubit velocities as a function of $\omega_B\beta_B$ with $\omega_A\beta_A=0.5$, and as a function of $\omega_A\beta_A$ with $\omega_B\beta_B=0.5$. The black dotted line corresponds to the static case $\upsilon_A\!=\!\upsilon_B\!=\!0$. Values below 2 indicate that the  bound is looser than the standard TUR.}
    \label{fig:ratio}
\end{figure}

In Fig. \ref{fig:bounds}, we present the fluctuations in heat exchanged with the hot bath, $\text{Var}[Q_H]$, for the case where either qubit A or B moves with constant relativistic velocity. The results are compared against the derived bound Eq. \eqref{bound:analytic}, the classical TUR bound Eq. \eqref{TUR}, as well as the generalized TUR bound introduced in \cite{TUR:SWAP}. The latter is given by
\begin{align}\label{TUR:gen}
    \frac{\text{Var}[\mathcal{J}]}{\expval{\mathcal{J}}^2}\geq f(\expval{\Sigma}),
\end{align}
where $f(x)=\text{csch}^2(g(x/2))$, with $\text{csch}(x)\equiv 1/\sinh x$, and $g(x)$ denoting the function inverse of $x\tanh(x)$. A Taylor expansion of $f(\expval{\Sigma})$ yields $f(\expval{\Sigma})\simeq2/\expval{\Sigma}-2/3$, and thus for $\expval{\Sigma}\ll 1$ the bound in Eq.~\eqref{TUR} is recovered \cite{TUR:SWAP}. We observe that the classical TUR bound can be violated for small values of the transition frequency ratio $\omega_B/\omega_A$, whereas both the bound Eq. \eqref{bound:analytic} and the generalized bound Eq. \eqref{TUR:gen} remain valid.

In Fig. \ref{fig:ratio}, we present the ratio
\begin{align}
\mathcal{R}\equiv\expval{\Sigma}\frac{\text{Var}[\mathcal{J}]}{\expval{\mathcal{J}}^2}
\end{align} for different qubit velocities $\upsilon_A$ and $\upsilon_B$, as a function of $\omega_B\beta_B$ with $\omega_A\beta_A=0.5$ fixed, and as a function of $\omega_A\beta_A$ with $\omega_B\beta_B=0.5$ fixed. This illustrates how the TUR bound is modified when the qubits are in motion compared to the rest-frame scenario ($\upsilon_A\!=\!\upsilon_B\!=\!0$). We observe that relativistic motion can lead to stronger violations of the bound already observed in the quantum regime. In particular, motion through the hot bath increases the violation when the machine operates as a refrigerator, whereas motion through the cold bath leads to a larger violation when the machine functions as a heat engine. These results suggest that relativistic motion may be harnessed as a resource to operate the machine beyond classical bounds (see also  Refs. \cite{DMOA2025surpassingcarnot,pandit2025, shastri2025relativisticmotionthermalbath}).

\section{Generalized performance bounds}\label{Sec:eff:bounds}

Now, we present performance bounds when the machine is functioning as a heat engine and, then as a refrigerator. The efficiency, $\eta$, of the heat engine mode of operation, defined as the ratio of the total work output to the absorbed heat, is given by
\begin{align}
     \eta=-\frac{ \expval{W}}{ \expval{Q_H}}=1-\frac{\omega_B}{\omega_A}.
\end{align}
This corresponds to the Otto efficiency, which depends only on the ratio of the transition frequencies of the two qubits. From the entropy production expression in Eq. \eqref{entropy}, and by invoking the second law of thermodynamics Eq. \eqref{2nd:law}, it follows that the maximum achievable efficiency in the engine regime (where $Q_H>0$) is
bounded by the generalized Carnot efficiency \cite{abah2014efficiency}
\begin{align}
     \eta\leq 1-\frac{T^{\text{eff}}_B}{T^{\text{eff}}_A} := \eta_C^{\text{eff}}.
     \label{efficiency}
\end{align}
In the static case, the standard Carnot limit is recovered.

Employing Eqs. \eqref{work:average}-\eqref{entropy}, we obtain a relation between the entropy production, the engine’s power  output $P=-\expval{W}$ and the generalized efficiency:
\begin{align}
    \expval{\Sigma}=\frac{\beta^{\text{eff}}_B\omega_B-\beta^{\text{eff}}_A\omega_A}{\omega_A-\omega_B}\expval{P}=\frac{\expval{P}}{T^{\text{eff}}_B}\left(\frac{\eta_C^{\text{eff}}}{\eta}-1\right).
\end{align}
The uncertainty relation Eq. \eqref{bound:analytic} then yields the following trade-off between the engine's efficiency and mean power output
\begin{align}
    \eta\leq\frac{\eta_C^{\text{eff}}}{1+2\expval{P}T^{\text{eff}}_B/\expval{P^2}}.
\end{align}
This implies that reaching efficiencies close to the generalized Carnot limit requires either a reduction in power output or an increase in the second moment of the work distribution. In our previous work \cite{DMOA2025surpassingcarnot}, we optimized the performance of the heat engine operating at finite power by evaluating its efficiency at maximum power, demonstrating that relativistic motion through the thermal field baths can yield an efficiency that exceeds the standard Carnot efficiency defined with respect to the rest-frame temperatures.

On the other hand, when the machine operates as a refrigerator, the first law of thermodynamics, $\expval{W}=-\expval{Q_H}-\expval{Q_C}$, can be used to determine its cooling power ($\expval{Q_C}>0$) as
\begin{align}\label{cooling:power}
    \expval{Q_C}&=\frac{\omega_B}{2}\bigg(\tanh(\beta^{\text{eff}}_A\omega_A/2)-\tanh(\beta^{\text{eff}}_B\omega_B/2)\bigg).
\end{align}
The coefficient of performance, $\varepsilon$, of the refrigerator,  defined as the ratio of heat absorbed from the cold environment to the total work input, is given by
\begin{align}
    \varepsilon=\frac{ \expval{Q_C}}{\expval{W}}=\frac{\omega_B}{\omega_A-\omega_B}.
\end{align}
Similarly to before, by combining the expression for entropy production in Eq. \eqref{entropy} together with the first law, we find that the second law imposes an upper bound on the performance of the refrigerator, given by the generalized COP 
\begin{align}\label{genCOP}  \varepsilon\leq\frac{1}{T^{\text{eff}}_A/T^{\text{eff}}_B-1}:=\varepsilon_C^{\text{eff}}.
\end{align}

We can also obtain a relation between the entropy production, the refrigerator's cooling power and the generalized COP as
\begin{align}
    \expval{\Sigma}=\frac{\beta^{\text{eff}}_A\omega_A-\beta^{\text{eff}}_B\omega_B}{\omega_B}\expval{Q_C}=\frac{\expval{Q_C}}{T^{\text{eff}}_A}\left(\frac{1}{\varepsilon}-\frac{1}{\varepsilon_C^{\text{eff}}}\right),
\end{align}
which implies the bound on the COP
\begin{align}\label{refrig:bound}
\varepsilon\leq\frac{\varepsilon_C^{\text{eff}}}{1+2T_A^{\text{eff}}\varepsilon_C^{\text{eff}}\expval{Q_C}/\expval{Q_C^2}}.
\end{align}
The figure of merit $\chi$ \cite{Abah_2016}, which quantifies the performance of a refrigerator per unit cycle time, is defined as the product of the COP and the cooling power, $\chi:=\varepsilon\expval{Q_C}$.
Consequently, Eq. \eqref{refrig:bound} indicates that enhancing the COP necessitates either a sacrifice in the figure of merit, which is analogous to the power output of a heat engine, or an increase of the second moment of the cooling power. For completeness, Appendix \ref{refrig:opt} presents an optimization of the refrigerator’s performance, showing that its COP may surpass the standard Carnot limit as a consequence of the effective temperatures perceived by the relativistically moving qubits.

%========================================
%========================================
\section{conclusions}\label{Sec:conclusions}

The study of thermodynamic uncertainty relations in relativistic settings remains largely unexplored. In this work, we employed a two-qubit SWAP engine whose working medium consists of inertially moving UDW qubit detectors to derive TURs that provide a trade-off between power output, fluctuations, entropy production, and efficiency in the presence of relative motion between the working medium and the thermal bath. Our results reveal that violations of the classical TUR—originally formulated for Markovian stochastic processes—arise not only from the quantum nature of the setup but also from relativistic motion. Furthermore, we derived generalized performance bounds for both the heat engine and refrigerator operating modes of relativistic quantum thermal machines, as constrained by the second law of stochastic thermodynamics. These results demonstrate that relativistic motion can enable operation beyond classical Carnot limits, offering new insights into the interplay between relativity, quantum effects, and nonequilibrium thermodynamics.

%========================================
%========================================
\section{acknowledgments}
This work was supported by the UK Research and Innovation Engineering and Physical Sciences Research Council (Grant No. EP/Z002796/1).
%===========================================================================================
\appendix
\section{Optimal performance of refrigerator}\label{refrig:opt}

We next focus on the high-temperature regime, where an inertially moving detector perceives an effective temperature, 
\begin{align}\label{teff:series}
    T^{\text{eff}}=\frac{T}{2\gamma\upsilon}\ln\left(\frac{1+\upsilon}{1-\upsilon}\right)+\mathcal{O}(\omega^2),
\end{align}
which is lower than the actual temperature of the thermal field bath.

\begin{figure}[bp]
    \centering    \includegraphics[width=0.52\linewidth]{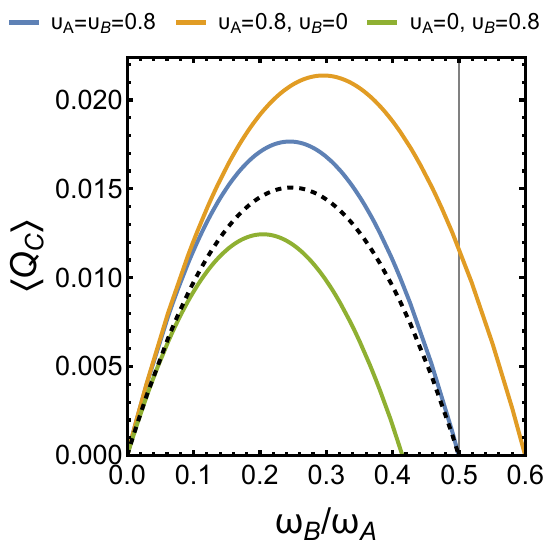}\hspace{0.18cm}\includegraphics[scale=0.42]{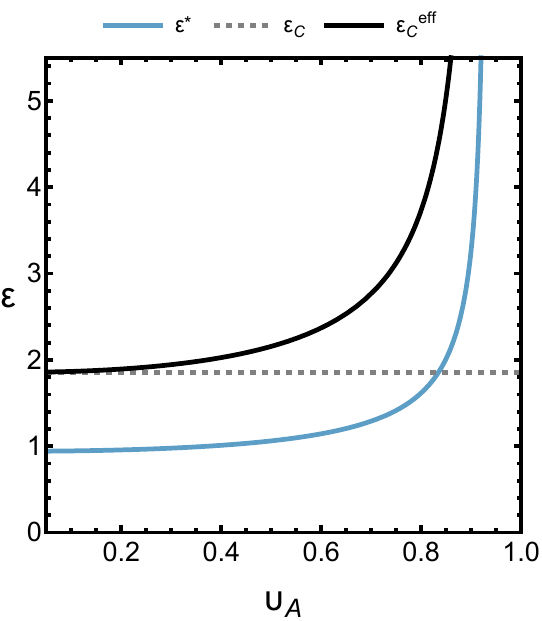}
    \caption{\justifying  \emph{Left panel}: Cooling power, $\expval{Q_C}$, as a function of the frequency ratio $\omega_B/\omega_A$, for a fixed temperature ratio of the thermal baths $\beta_A/\beta_B = 1/2$ and varying qubit speeds. The black dotted line corresponds to the static case $\upsilon_A\!=\!\upsilon_B\!=\!0$. The vertical line indicates the boundary between refrigerator and heat engine operational regimes in the rest-frame case. \emph{Right panel}: COP at maximum figure of merit, $\varepsilon^*$, as a function of the velocity of the qubit $A$, when  $\beta_A/\beta_B = 0.65$. Here, $\varepsilon_C$ is the standard Carnot COP, and $\varepsilon_C^{\text{eff}}$ the generalized Carnot bound. }
    \label{fig:cooling:power}
\end{figure}

In Fig. \ref{fig:cooling:power}, we present the average cooling power $\expval{Q_C}$ as a function of the frequency ratio $\omega_B/\omega_A$ for a fixed temperature
ratio $\beta_A/\beta_B=1/2$. We consider the cases where both qubits A and B move through their respective thermal reservoirs at the same relativistic velocity, $\upsilon_A=\upsilon_B$; only qubit A moves through the hot bath; and only qubit B moves through the cold bath. For comparison, we also include the case in which both qubits remain at rest, corresponding to the standard quantum SWAP thermal machine. We observe that the device can operate as a refrigerator in parameter regimes where it would act as a heat engine if both qubits were static. In particular, the device functions as a refrigerator when $\omega_B/\omega_A<\beta^{\text{eff}}_A/\beta^{\text{eff}}_B$. For instance, when qubit $A$ moves at speed $\upsilon_A=0.8$, while qubit $B$ remains at rest,  Eq. \eqref{teff:series} indicates that the device operates as a refrigerator for $\omega_B/\omega_A\lesssim 0.6$, in agreement with the results shown in the figure. Moreover, the cooling power increases when qubit $A$ moves at relativistic speed through the hot bath. The cooling power also appears to increase when both qubits move at the same velocity; however, in this case, Eq. \eqref{genCOP} shows that the refrigerator’s COP remains bounded by the classical Carnot limit, $\varepsilon_C$. 
%===========================================================================================================================
%===========================================================================================================================

We analyze the refrigerator's performance by evaluating its COP at maximum figure of merit $\chi$. In particular, we optimize its performance with respect to the qubit frequency $\omega_B$, by solving the equation $\partial\chi/\partial\omega_B=0$. The figure of merit is maximized when the qubit frequencies satisfy
\begin{align}
\frac{\omega_B}{\omega_A}\!=\frac{\left(\!\beta_A^{\text{eff}}\!+\!3\beta_B^{\text{eff}}\!-\!\sqrt{(\beta_A^{\text{eff}})^2\!-\!10\beta_A^{\text{eff}}\beta_B^{\text{eff}}\!+\!9(\beta_B^{\text{eff}})^2}\!\right)}{4\beta_B^{\text{eff}}}.
\end{align}
The corresponding COP at maximum figure of merit is then given by
\begin{align}\label{COPopt}
\varepsilon^*=\frac{1}{2}\left(\sqrt{8\varepsilon_C^{\text{eff}}+9}-3\right).
\end{align}
In the static limit, this reduces to the COP at the maximum figure of merit for a two-level quantum Otto refrigerator \cite{singh:OA}.
We plot the optimized COP \eqref{COPopt} in Fig. \ref{fig:cooling:power} as a function of the velocity of the qubit $A$, while keeping qubit $B$ static. We find that the refrigerator can achieve a COP that can exceeds the Carnot limit $\varepsilon_C$---a behavior reminiscent of the enhancement observed in refrigerators where the working medium is coupled to squeezed thermal reservoirs \cite{Long:Wei}.

\bibliography{references}

\end{document}